\documentclass[11pt,twoside]{article}
\usepackage{./asp2014}
\usepackage{hyperref}
\aspSuppressVolSlug
\resetcounters
\begin{document}

\title{Detrending algorithms in large time-series: Application to TFRM-PSES data}
\author{{\small del Ser D.,$^{1,3}$ Fors O.,$^{2,3}$ N\'{u}\~{n}ez J.,$^{1,3}$ Voss H.,$^3$ Rosich A.$^{1,4}$ and Kouprianov V.$^5$}
\affil{{\small $^1$Reial Academia de Ciencies i Arts de Barcelona (RACAB), Barcelona, Spain;}}
\affil{{\small $^2$Department of Physics and Astronomy, University of North Carolina at Chapel Hill, Chapel Hill, NC, USA;}}
\affil{{\small $^3$Dept. d'Astronomia i Meteorologia and Institut de Ciències del Cosmos (ICC), Universitat de Barcelona (UB), Barcelona, Spain;}}
\affil{{\small $^4$Institut de Ciencies de l'Espai ICE (CSIC-IEEC);}}
\affil{{\small $^5$Central (Pulkovo) Astronomical Observatory of Russian Academy of Sciences, St. Petersburg, Russia;}}
\email{dser@am.ub.es}}

\paperauthor{Daniel del Ser}{dser@am.ub.es}{}{Reial Academia de Ciencies i Arts de Barcelona (RACAB)}{Dept. d'Astronomia i Meteorologia, Universitat de Barcelona}{Barcelona}{Barcelona}{08028}{Spain}
\paperauthor{Octavi Fors}{octavi@live.unc.edu}{}{University of North Carolina at Chapel Hill}{Department of Physics and Astronomy}{Chapel Hill}{NC}{27599-3255}{USA}
\paperauthor{Jorge N\'{u}\~{n}ez}{jorge@am.ub.es}{}{Reial Academia de Ciencies i Arts de Barcelona (RACAB)}{Dept. d'Astronomia i Meteorologia and Institut de Ciències del Cosmos (ICC)}{Barcelona}{Barcelona}{08028}{Spain}
\paperauthor{Holger Voss}{hvoss@am.ub.es}{}{Universitat de Barcelona}{Dept. d'Astronomia i Meteorologia and Institut de Ciències del Cosmos (ICC)}{Barcelona}{Barcelona}{08028}{Spain}
\paperauthor{Albert Rosich}{albertrosich@gmail.com}{}{Institut de Ciencies de l'Espai ICE (CSIC-IEEC)}{}{Barcelona}{Barcelona}{}{Spain}
\paperauthor{Vladimir Kouprianov}{v.k@bk.ru}{}{Central (Pulkovo) Astronomical Observatory of Russian Academy of Sciences}{}{St. Petersburg}{}{}{Russia}

\begin{abstract}
{\small Certain instrumental effects and data reduction anomalies introduce systematic errors in photometric time-series. Detrending algorithms such as the Trend Filtering Algorithm (TFA) \citep{ex_2} have played a key role in minimizing the effects caused by these systematics. Here we present the results obtained after applying the TFA, Savitszky-Golay \citep{ex_3} detrending algorithms and the Box Least Square phase folding algorithm \citep{ex_1} to the TFRM-PSES data \citep{ex_4}. Tests performed on this data show that by applying these two filtering methods together, the photometric RMS is on average improved by a factor of 3-4, with better efficiency towards brighter magnitudes, while applying TFA alone yields an improvement of a factor 1-2. As a result of this improvement, we are able to detect and analyze a large number of stars per TFRM-PSES field which present some kind of variability. Also, after porting these algorithms to Python and parallelizing them, we have improved, even for large data samples, the computing performance of the overall detrending+BLS algorithm by a factor of $\sim$10 with respect to \citet{ex_2}.}
\end{abstract}
\vspace{-1.2cm}
\section{Introduction}
{\small In this paper we present the results obtained after applying the combination of two known detrending algorithms (Kov\'{a}cs TFA and the Savitszky-Golay filter) to 1380 high-quality images coming from the TFRM-PSES field 964622 comprising 41 nights from the 29 May 2012 to the 13 Aug 2014.\\The TFRM project consists in the refurbishment of the f/1 50 cm ROA's Baker-Nunn Camera for robotic CCD surveying purposes \citep{ex_4}. Currently installed at the summit of the Montsec d'Ares (Lleida), is a joint collaboration between the Reial Acad\`{e}mia de Ci\`{e}ncies i Arts de Barcelona (RACAB), the Real Instituto y Observatorio de la Armada (ROA) with the participation of several members of the Departament d'Astronomia i Meteorologia, Universitat de Barcelona. The \href{http://www.am.ub.edu/bnc}{TFRM-PSES} survey is an ongoing systematic search for super-Earths orbiting M-dwarf stars \citep{ex_4}. The TFRM-PSES, with sufficient cadence, is able to monitor multiple fields observed every night, each one containing $\sim$20 M-type catalogued stars, in the range of 9.0mag < V < 15.5mag.}

\section{Detrending Algorithms}

{\small The number of high precision light curves that is obtainable with a photometric survey system depends on several factors such as the atmosphere, the instrument or the photometric errors of the measurements. These systematic and random noise sources create trends in our time-series that undermine the intrinsic signals of stars.
TFA removes trends from trend- and noise-dominated time series and reconstructs the shape of periodic signals taking into account the fact that many of the systematic variations in a given light curve are shared by other stars in the same data set. Using a sample of reference stars, one can apply some kind of optimum filtering (in our case, a linear combination) to the data.
A Savitzky-Golay filter is a digital filter used to smooth a set of data points without greatly distorting the signal. This is achieved, through convolution, by fitting successive sub-sets of adjacent data points with a low-degree polynomial by the linear least square method. This algorithm tends to keep characteristics of the original data point distribution such as the relative maxima and minima and the width of the peaks. These two algorithms have been exported to Python, combined and parallelized.}
\section{TFRM-PSES data results}

{\small Photometric analysis is based on the principles of the BESTRED pipeline \citep{ex_5}. We use two python-based scripts, inspired on the previous algorithm and encompassed in the APEX pipeline developed by Kouprianov V.
Parallelization of the detrending algorithms has allowed us to decrease the computation time to less than 3CPU hours for almost 3000 stars with at least 1000 points and $\sim$200 template stars (last runs on the script to $\sim$16000 stars yield less than 30 hours of CPU time) a factor 10 less with respect to \citet{ex_2}.
After applying both algorithms, overall and intra-night dipersion decreases thus improving RMS from $\sim$0.3mag to $\sim$2-3mmag for the brightest, non-saturated stars and getting closer to the 'theoretical' statistical error computed from the quadratic sum of the background, photon and scintillation noises (see Figure \ref{ex_fig1}).}
\vspace{-0.2cm}
\articlefigure[scale=.3]{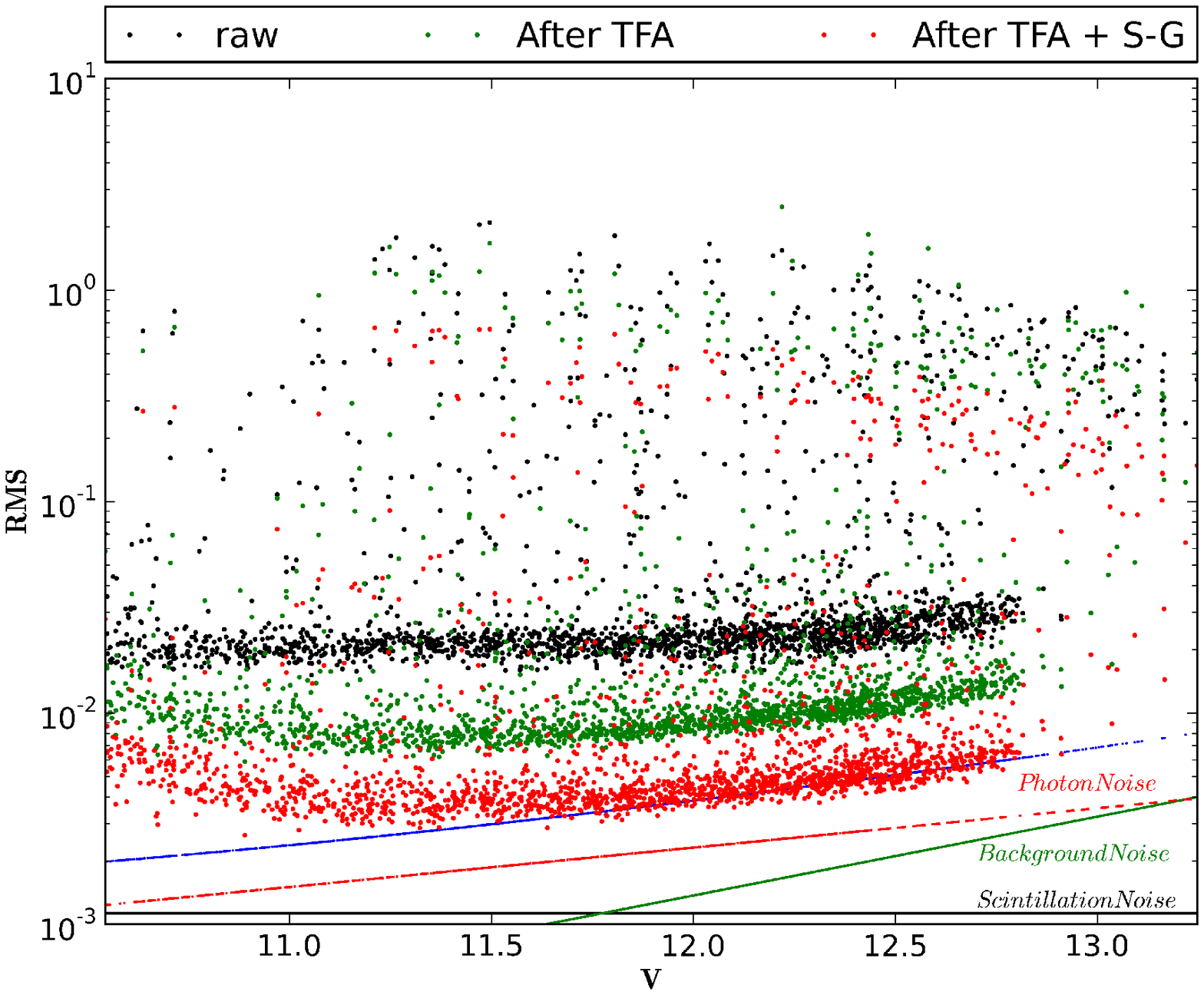}{ex_fig1}{\small{RMS vs V plot for PSES field 964622.}}


\begin{thebibliography}{}
\bibitem[Kov\'{a}cs et al., 2002]{ex_1}
Kov\'{a}cs et al., 2002. A\&A, 391
\bibitem[Kov\'{a}cs et al., 2004]{ex_2}
Kov\'{a}cs et al., 2005. MNRAS, 356
\bibitem[Savitzky \& Golay, 1964]{ex_3}
Savitzky, A. \& Golay, M.J.E., 1964. Analytical Chemistry, 36
\bibitem[Fors et al., 2013]{ex_4}
Fors et al., 2013. PASP, 125
\bibitem[Voss, 2006]{ex_5}
Voss, H., 2006. Ph.D. Thesis, University of Berlin.
\end{thebibliography}
\vspace{-0.5cm}

\end{document}